\documentclass{mn2e}
\input{epsf}

\title[Dark matter distribution in Draco]{Dark matter distribution in
the Draco dwarf from velocity moments}

\author[E. L. {\L}okas, G. A. Mamon and F. Prada]
    {Ewa L. {\L}okas,$^{1}$\thanks{E-mail: lokas@camk.edu.pl}
    Gary A. Mamon$^{2,3}$\thanks{E-mail: gam@iap.fr} and
    Francisco Prada$^{4}$\thanks{E-mail: fprada@iaa.es}
    \\
    $^1$Nicolaus Copernicus Astronomical Center, Bartycka 18,
    00-716 Warsaw, Poland\\
    $^2$Institut d'Astrophysique de Paris (UMR 7095: CNRS \& Universit\'e
    Pierre \& Marie Curie), 98 bis Bd Arago,
    F-75014 Paris, France\\
    $^3$GEPI (UMR 8111: CNRS \& Universit\'e Denis Diderot), Observatoire de Paris, F-92195 Meudon, France\\
    $^4$Instituto de Astrof{\'\i}sica de Andalucia (CSIC),
    Apartado Correos 3005, E-18080 Granada, Spain}
\begin{document}

\maketitle

\begin{abstract}
We study the distribution of dark matter in the Draco dwarf spheroidal galaxy by
modelling the moments of the line-of-sight velocity distribution of stars obtained from
new velocity data of Wilkinson et al. The luminosity distribution is approximated
by a S\'ersic profile fitted to the data
by Odenkirchen et al. We assume that the dark matter density
profile is given by a formula with an inner cusp and an outer exponential
cut-off, as
recently proposed by Kazantzidis et al. as a result of simulations of
tidal stripping of dwarfs by the potential of the Milky Way.
The dark matter distribution is characterized by
the total dark mass and the cut-off radius.
The models have arbitrary velocity anisotropy parameter assumed to be constant with radius.
We estimate the three parameters by fitting
both the line-of-sight velocity dispersion and kurtosis profiles, which
allows us to break the degeneracy between the mass distribution and velocity
anisotropy. The results of the fitting procedure turn out to be very different
depending on the stellar sample considered, that is on our choice of stars
with discrepant velocities to be discarded as interlopers.
For our most reliable sample,
the model parameters remain weakly constrained, but the robust result is the preference
for weakly tangential stellar orbits and high mass-to-light ratios.
The best-fitting total mass is then $7 \times 10^7 M_{\sun}$,
much lower than recent estimates, while the mass-to-light ratio is $ M/L_{\rm V} = 300$
and almost constant with radius. If the binary fraction in the stellar population of Draco
turns out to be significant, the kurtosis of the global velocity distribution
will be smaller and the orbits inferred
will be more tangential, while the resulting mass estimate lower.

\end{abstract}


\section{Introduction}

Dwarf spheroidal galaxies are currently believed to be among the most dark matter
dominated objects in the Universe (with mass-to-light ratios up to few hundred solar units)
and therefore are of immediate interest when
trying to test present theoretical predictions concerning the dark matter profiles.
As dwarfs, they are also important for theories of structure formation, which suffer
from the overabundance of subhaloes (Klypin et al. 1999).

The Draco dwarf spheroidal galaxy (hereafter, Draco) is a generic example of this class of objects
in the neighbourhood of the Galaxy and has been already the subject of extensive study.
Recently, new radial-velocity observations of Draco have yielded a larger kinematic sample
(Kleyna et al. 2002; Wilkinson et al. 2004) and the Sloan Digital Sky Survey has
yielded a better determination of the luminosity profile and shape (Odenkirchen et al. 2001).
Draco has also been examined for the presence of tidal tails which could indicate
a kinematically perturbed state.
For a long time it has been debated (e.g. Klessen \& Kroupa 1998;
Klessen \& Zhao 2002) whether the large velocity dispersion measured
in Draco is a result of significant dark matter content or a signature of tidal
disruption by the Milky Way. New analyses by Piatek et al. (2002) and
Klessen, Grebel \& Harbeck (2003) show that there is no evidence that tidal
forces from the Milky Way have disturbed the inner structure of Draco.
In particular, Klessen et al. (2003) argue that models without any
dark matter are unable to reproduce the narrow observed horizontal branch width of Draco.
As we will show below, the truth may lie between these two pictures:
Draco most probably is strongly dark matter
dominated but still its outer edges could be affected by tidal interaction with the Milky
Way producing a population of stars unbound to the dwarf.

One of the major methods to determine the dark matter content of galaxies and clusters
is to study the kinematics of their observable discrete component: stars or galaxies, respectively.
The method is based on the assumption that the positions and velocities of the
component members
trace the Newtonian gravitational potential of the object.
If the object is in virial equilibrium
the relation between those is described by the Jeans formalism. In the
classical approach, such an analysis is restricted to solving the lowest order Jeans
equation and modelling only the velocity
dispersion profile. In previous papers (\L okas 2002; \L okas \& Mamon 2003),
building on the earlier work by Merrifield \& Kent (1990) and van der Marel et al.
(2000), we have extended the formalism to include the solutions of the higher-order
Jeans equations describing the fourth velocity moment, the kurtosis.

The formalism
has been successfully applied to study the dark matter distribution in the Coma cluster
of galaxies (\L okas \& Mamon 2003). We have shown that, for a restricted class of
dark matter distributions motivated by the results of cosmological $N$-body simulations,
the joint analysis of velocity
dispersion and kurtosis allows us to break the usual degeneracy between the mass
distribution and velocity anisotropy and constrain the parameters of the dark matter
profile. Recently we have tested the reliability of this approach against a series
of $N$-body simulations (Sanchis, \L okas \& Mamon 2004) and verified that the
method allows for typically accurate determinations of the mass, concentration and
velocity anisotropy (albeit with a large scatter in concentration).
In the present paper, encouraged by these successes, we apply the method
to constrain the dark matter distribution in the Draco dwarf spheroidal
galaxy.

The paper is organized as follows. In Section~2 we summarize the Jeans formalism for the
calculation of the moments of the line-of-sight velocity distribution. In Section~3
we estimate the moments from the velocity data and discuss the associated difficulties
caused by the possible presence of interlopers and binary stars.
Section~4 describes our models for the luminosity and dark matter distributions. Results
of fitting the models to the data are presented and discussed in Section~5. In Section~6
we comment on the possible origin of unbound stars in Draco and the concluding remarks
follow in Section~7.

\section{Velocity moments from the Jeans formalism}

The Jeans formalism (e.g. Binney \& Tremaine 1987) relates the velocity moments
of a gravitationally bound object to the underlying mass distribution. We summarize
here the formalism, as developed in \L okas (2002) and \L okas \& Mamon (2003).
The second $\sigma_r^2$ and fourth-order $\overline{v_r^4}$ radial velocity moments
obey the Jeans equations
\begin{eqnarray}    \label{m1}
	\frac{\rm d}{{\rm d} r}  (\nu \sigma_r^2) + \frac{2 \beta}{r} \nu
	\sigma_r^2 + \nu \frac{{\rm d} \Phi}{{\rm d} r} &=& 0   \label{jeans1} \\
 	\frac{\rm d}{{\rm d} r}  (\nu \overline{v_r^4}) + \frac{2 \beta}{r} \nu
	\overline{v_r^4} + 3 \nu \sigma_r^2 \frac{{\rm d} \Phi}{{\rm d} r} &=& 0
	\label{jeans2}
\end{eqnarray}
where $\nu$ is the 3D density distribution of the tracer population and $\Phi$ is
the gravitational potential. The second equation was derived assuming the distribution
function of the form $f(E,L)=f_0(E) L^{-2 \beta}$ and
the anisotropy parameter relating the angular $\sigma_\theta$ and radial
$\sigma_r$ velocity dispersions
\begin{equation}    \label{beta}
	\beta=1-\frac{\sigma_\theta^2(r)}{\sigma_r^2(r)}
\end{equation}
to be constant with radius. We will consider here $-\infty < \beta \le 1$ which covers
all interesting possibilities from radial orbits
($\beta=1$) to isotropy ($\beta=0$) and circular orbits
($\beta \rightarrow - \infty$).

The solutions to equations (\ref{jeans1})-(\ref{jeans2}) are
\begin{eqnarray}
	\nu \sigma_r^2 (\beta={\rm const}) &=& r^{-2 \beta}
	\int_r^\infty r^{2 \beta} \nu \frac{{\rm d} \Phi}{{\rm d} r} \ {\rm d}r
	\label{sol1} \\
	\nu \overline{v_r^4} (\beta={\rm const}) &=& 3 r^{-2 \beta}
	\int_r^\infty r^{2 \beta} \nu \sigma_r^2 (r)
	\frac{{\rm d} \Phi}{{\rm d} r} \ {\rm d} r \ .
	\label{sol2}
\end{eqnarray}
Projecting along the line of sight we obtain the observable quantities
\begin{eqnarray}
        \sigma_{\rm los}^2 (R) &=& \frac{2}{I(R)} \int_{R}^{\infty}
	\frac{\nu \sigma_r^2 r}{\sqrt{r^2 - R^2}} \left( 1-\beta \frac{R^2}{r^2} \right)
	{\rm d} r
	\label{proj1} \\
	\overline{v_{\rm los}^4} (R) &=& \frac{2}{I(R)} \int_{R}^{\infty}
	\frac{\nu \,  \overline{v_r^4} \,r}{\sqrt{r^2 - R^2}} \ g(r, R, \beta)
	\,{\rm d} r
	\label{proj2}
\end{eqnarray}
where
\begin{equation}	\label{proj2a}
    g(r, R, \beta) = 1 - 2 \beta \frac{R^2}{r^2} + \frac{\beta(1+\beta)}{2}
    \frac{R^4}{r^4} \ ,
\end{equation}
$I(R)$ is the surface distribution of the tracer and
$R$ is the projected radius.

Introducing equations (\ref{sol1})-(\ref{sol2}) into (\ref{proj1})-(\ref{proj2})
and inverting the order of integration, the calculations of $\sigma_{\rm los}$
and $\overline{v_{\rm los}^4}$ can be reduced to one- and two-dimensional numerical
integrals respectively (see the appendix of Mamon \& {\L}okas 2005 for a simpler
expression for $\sigma_{\rm los}$). In the following we will refer to the fourth moment in the
form of kurtosis
\begin{equation}	\label{kurt}
	\kappa_{\rm los} (R) = \frac{\overline{v_{\rm los}^4} (R)}
	{\sigma_{\rm los}^4 (R)} \ ,
\end{equation}
whose value is 3 for a Gaussian distribution.

\section{Velocity moments from observations}

\begin{figure}
\begin{center}
    \leavevmode
    \epsfxsize=8cm
    \epsfbox[40 40 320 300]{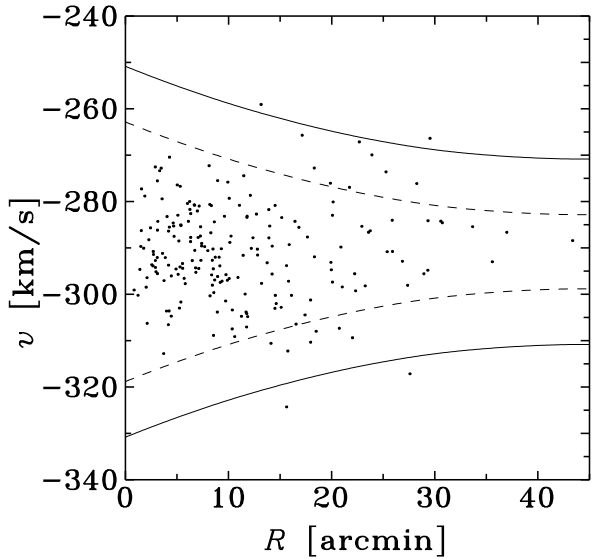}
\end{center}
\caption{Heliocentric velocities versus distances from the centre of Draco for
207 stars. The solid (dashed) curves separate 203 (189) stars
counted as members of the galaxy from the 4 (18) supposed interlopers. }
\label{stars}
\end{figure}

Fig.~\ref{stars} shows the heliocentric velocities and projected distances from
the centre of 207 stars with good velocity measurements from Wilkinson et al. (2004).
The distances are
calculated assuming that the centre of Draco is at RA=$17^{\rm h}20^{\rm m}13.2^{\rm s}$,
Dec=$57^\circ54'54''$ (J2000) (Odenkirchen et al. 2001).
The sample was obtained from the original sample of 416 velocities stars observed in
the direction of Draco by cutting out all objects with velocities differing by more
than 39 km s$^{-1}$ from Draco's mean velocity of $-291$ km s$^{-1}$.
This choice was motivated by the assumption that the velocity distribution is Gaussian
with a maximum dispersion of 13 km s$^{-1}$. We should however still
take into account the possibility that some of the 207 stars are interlopers unbound
to the gravitational potential of the galaxy. Prada et al. (2003) discussed the
issue thoroughly in the case of isolated galaxies in the Sloan Digital Sky Survey
(SDSS) and showed that the presence of interlopers can affect significantly
the velocity dispersion and therefore the inferred mass distribution in the galaxy.
This effect is especially important in the case of small samples like the present one.

The problem has
been also discussed by Mayer et al. (2001) who showed (see Fig. 26 of the published version
of their paper) how the velocity dispersion profiles
of the dwarf spheroidal galaxies can be affected if the line of sight is along one of the
tidal tails: while the intrinsic dispersion profile is declining, the observed one, due
to the presence of unbound stars in the tail, is much higher and increasing. They state
explicitly that a velocity dispersion that increases outside the core of the dwarf
should be considered as a result of possible contamination by tidal tails. We address the
issue in more detail in Section~6.

Judging by the shape of the diagram shown in Fig.~\ref{stars} and
comparing it with the expected appearance of gravitationally bound object in velocity
space we immediately see that there are at least
four stars with discrepant velocities: they
have been separated from the main body of the galaxy by the solid lines. An even more
stringent, but still arbitrary choice, with 14 more stars excluded, is shown
with dashed lines. The lines were drawn symmetrically with respect to
the mean systemic velocity of $-291$ km s$^{-1}$, as estimated from the total sample
with 207 stars.
Since we have no possibility of determining a priori which stars are actually bound to
the galaxy we calculate the velocity moments using the three samples with 207, 203 and
189 velocities obtained in this way.

In the estimation of the moments, we follow the Appendix in \L okas \& Mamon (2003)
neglecting the small errors in the velocity measurements.
We divide the sample into 5 radial bins each containing 37-43 stars (depending on the
sample: 207 stars are divided into bins with $4 \times 41 + 43$ velocities, 203 stars into
$4 \times 41 + 39$ velocities and 189 stars into $4 \times 38 + 37$ velocities).
The most natural estimators of the variance and kurtosis from a sample of
$n$ line-of-sight velocity measurements $v_i$ are
\begin{equation}    \label{app1}
	S^2 = \frac{1}{n} \sum_{i=1}^n (v_i - \overline{v})^2
\end{equation}
and
\begin{equation}    \label{app2}
	K = \frac{\frac{1}{n} \sum_{i=1}^n (v_i - \overline{v})^4}{(S^2)^2}
\end{equation}
where
\begin{equation}    \label{app3}
	\overline{v} = \frac{1}{n} \sum_{i=1}^n v_i
\end{equation}
is the mean of stellar velocities in the sample.
The distributions of these estimators for our binning of stars,
i.e. when $n \approx 40$, can be investigated by running Monte Carlo simulations
selecting ${\cal N}=10^4$ times
$n=40$ numbers from a Gaussian distribution with zero mean and dispersion of unity
(see \L okas \& Mamon 2003).
One can then construct unbiased and nearly Gaussian-distributed estimators of line-of-sight
velocity dispersion $s$ and kurtosis-like variable $k$
\begin{equation}    \label{app4}
	s = \left( \frac{n}{n-1} S^2 \right)^{1/2}
\end{equation}
\begin{equation}    \label{app5}
	k = \left[ \log \left( \frac{3}{2.75} K \right) \right]^{1/10} .
\end{equation}
The factor $n-1$ in equation (\ref{app4}) is the well known correction for bias when
estimating the sample variance, valid independently of the underlying distribution.
In (\ref{app5}) the factor $3/2.75$ corrects for the bias in the kurtosis estimate,
i.e. unbiased estimate of kurtosis is $K' = 3K/2.75$, while
the rather complicated function of $K'$ assures that the sampling distribution of $k$ is
approximately Gaussian. We find that the standard errors
in the case of $s$ are of the order of 11 percent
while in the case of $k$ are approximately 2 percent. In the following we
assign these errors to our data points.

We checked that even for weakly non-Gaussian velocity distributions, such as following from
the present data, estimators $s$ and $k$ are Gaussian-distributed to a very good
approximation and very weakly correlated with the correlation coefficient
$|\varrho| \le 0.07$. We can therefore
assume that, to a good approximation, all of our data points measuring
velocity dispersion and kurtosis will be independent, which justifies the use
of standard $\chi^2$ minimization to fit the models to the data.

Apart from the sampling distributions discussed above, our estimates of the velocity
moments can be affected by the presence of binary stars.
Contrary to the studies of galaxy clusters, where galaxy pairs are rare and can be rather easily
identified and eliminated from the sample, in the case of stellar systems, such as Draco,
binaries can be identified only through long-term monitoring of the whole stellar
sample. In the sample of Wilkinson et al. (2004) such repeated measurements exist only for
66 stars and among those only a few are probable binaries, i.e. the differences between
their velocity measurements are too big to be accounted for by the measurement errors.
Given the uncertainties in our knowledge of the binary population in Draco, we assume
that the binary fraction is negligible ($\gamma =0$), but we discuss briefly below how a
significant amount of binaries could affect the results.

De Rijcke \& Dejonghe (2002) studied the influence of the binary population on the
observed line-of-sight velocity moments and found (see their Fig. 15)
that for stellar systems with
$\sigma_{\rm los} \approx 9$ km s$^{-1}$, such as Draco, the observed (Gaussian) dispersion
is mildly affected by the presence of binaries in
comparison with the intrinsic stellar velocity dispersion. However,
the kurtosis value turns out to be more affected.
Inverting the formulae (55) in De Rijcke \& Dejonghe (2002) we find that given the
line-of-sight velocity moments of the binary population, $\sigma_{\rm b}$ and
$\kappa_{\rm b}$ we can calculate the intrinsic moments of the stars $\sigma_{\rm i}$ and
$\kappa_{\rm i}$ from the observed ones $\sigma_{\rm o}$ and
$\kappa_{\rm o}$
\begin{eqnarray}
	\sigma_{\rm i}^2 &=& \sigma_{\rm o}^2 - \gamma \sigma_{\rm b}^2
	\label{bin1} \\
	\kappa_{\rm i} &=& \kappa_{\rm o} \frac{\sigma_{\rm o}^4}{\sigma_{\rm i}^4}
	- \gamma \kappa_{\rm b} \frac{\sigma_{\rm b}^4}{\sigma_{\rm i}^4}
	-6 \gamma \frac{\sigma_{\rm b}^2}{\sigma_{\rm i}^2} \ .
	\label{bin2}
\end{eqnarray}
We adopt the values of $\sigma_{\rm b} = 2.87$ km s$^{-1}$ and
$\kappa_{\rm b} =86.89$, as obtained by De Rijcke \& Dejonghe (2002) for their standard
binary population model. Having calculated the observed moments $\sigma_{\rm o}=s$
and $\kappa_{\rm o}= K'= 3K/2.75$
from equations (\ref{app1})-(\ref{app4}) we obtain the intrinsic values from equations
(\ref{bin1})-(\ref{bin2}).

\begin{figure}
\begin{center}
    \leavevmode
    \epsfxsize=8cm
    \epsfbox[40 40 320 300]{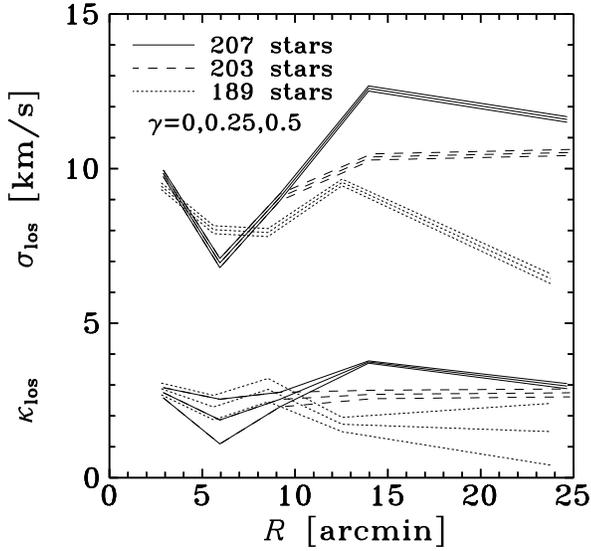}
\end{center}
\caption{Velocity moments calculated from the three samples of 207 (solid lines),
203 (dashed lines) and 189 stars (dotted lines) and different binary fraction
$\gamma=0$, 0.25 and 0.5. In each triplet the highest line
corresponds to $\gamma=0$ (no correction for binaries).}
\label{dispkurt}
\end{figure}

The uncertainties in the determination of the velocity moments are illustrated in
Fig.~\ref{dispkurt}. The solid lines join the moments calculated in five bins
with all 207
velocities, the dashed and dotted ones with 203 and 189 velocities respectively.
The three curves of each kind correspond to the binary fraction $\gamma=0$, 0.25 and 0.5,
from the highest to the lowest, i.e. correcting for binaries decreases the values of
moments. The estimates overlap for the first three bins of samples with 207 and 203 stars
because none of the
supposed interlopers fall there, the small differences in the case of the sample with
189 stars are due to different number
of stars per bin. For the two outer bins the situation is however very different,
with all 207 stars the velocity dispersion is much larger than for
203 and 189 stars. We note that although the profile for 207 stars decreases in the last
bin, there is no sudden drop in velocity dispersion as found by Wilkinson et al. (2004).
This is mainly due to our larger number of stars per bin.

As demonstrated by Fig.~\ref{dispkurt}, except for the most numerous sample with 207 stars,
the kurtosis values typically fall below the Gaussian value of 3.
Such kurtosis signifies a centrally
flattened velocity distribution in comparison with the Gaussian.
It is worth noting that the behaviour of the
kurtosis profile for the sample with 189 stars
(close to Gaussian value in the centre and lower values at larger
distances) is similar to the one found in the case of distribution of galaxy
velocities in the Coma cluster (\L okas \& Mamon 2003).
Decreasing kurtosis values seem also to be typical properties of the projected velocity
distributions of particles in simulated dark matter haloes
as demonstrated by the studies of Sanchis et al. (2004), Kazantzidis,
Magorrian \& Moore (2004) and Diemand, Moore \& Stadel (2004).

Our purpose now is to reproduce the observed profiles of velocity moments using models
described in the next Section.

\section{Distributions of stars and dark matter}

\subsection{Stars}

The distribution of stars is modelled in the same way as in \L okas (2001, 2002)
i.e. by the S\'ersic profile (S\'ersic 1968)
\begin{equation}    \label{m5}
	I(R) = I_0 \exp [-(R/R_{\rm S})^{1/m}],
\end{equation}
where $I_0$ is the central surface brightness and $R_{\rm S}$ is the
characteristic projected radius of the S\'ersic profile.
The 3D luminosity density $\nu(r)$ is obtained from $I(R)$ by
deprojection
\begin{equation}	\label{m5a}
	\nu(r) = - \frac{1}{\pi} \int_r^\infty \frac{{\rm d} I}{{\rm d} R}
	\frac{{\rm d} R}{\sqrt{R^2 - r^2}}.
\end{equation}

For a wide range of values $1/2 \le m \le 10$ found  to fit the light distribution
in elliptical and spheroidal galaxies an excellent
approximation for $\nu(r)$ is provided by (Lima Neto, Gerbal \&
M\'arquez 1999)
\begin{eqnarray}
	\nu(r) &=& \nu_0 \left( \frac{r}{R_{\rm S}} \right)^{-p}
	\exp \left[-\left( \frac{r}{R_{\rm S}}
	\right)^{1/m} \right]
	\label{m5c} \\
	\nu_0 &=& \frac{I_0 \Gamma(2 m)}{2 R_{\rm S}
	\Gamma[(3-p) m]}  \nonumber \\
	p &=& 1.0 - 0.6097/m + 0.05463/m^2 \nonumber.
\end{eqnarray}
The mass distribution of stars following from (\ref{m5c}) is
\begin{equation}	\label{m5d}
	M_{*}(r) = M_{\rm S} \frac{\gamma[(3-p)m,
	(r/R_{\rm S})^{1/m}]}{\Gamma[(3-p)m]},
\end{equation}
where $M_{\rm S} = \Upsilon L_{\rm tot}$ is the total mass of stars, with
$\Upsilon$=const the mass-to-light ratio for stars, $L_{\rm tot}$
the total luminosity of the galaxy, and $\gamma(\alpha, x) = \int_0^x
{\rm e}^{-t} t^{\alpha-1} {\rm d} t$ is the incomplete gamma function.

Odenkirchen et al. (2001) find that their data for the surface density of stars in Draco
are best fitted (for sample S1) with $1/m=1.2$ ($m=0.83$) and the S\'ersic radius
$R_{\rm S}=7.3$ arcmin. We will adopt these values here assuming
that the population of stars is uniform
and therefore the distribution of stars can
be translated to the surface luminosity in any band just by adjusting
the normalization $I_0$. We note that the surface
density distribution estimated by Odenkirchen et al. (2001) agrees well with the
independent estimates by Piatek et al. (2002) from a different data set.
In particular, both groups find
the distribution to fall sharply at large distances thus contradicting the earlier
estimates of Irwin \& Hatzidimitriou (1995) who found a flattening profile which could
point towards the presence of tidal tails.

We adopt the apparent magnitude of Draco in the $i$-band $m_i = 10.49$ following
Odenkirchen et al. (2001). Using their formula (8) relating $i$ and $I$ magnitudes,
and assuming $V-I =1$ from the colour-magnitude diagram in Fig. 5 of Bonanos et al.
(2004) we find $i-I = 0.52$ in agreement with Table~3 of Fukugita, Shimasaku \&
Ichikawa (1995) and $V-i=0.48$. Taking a distance estimate for Draco of $d=80$ kpc
(Aparicio, Carrera \& Martinez-Delgado 2001), corresponding to the
distance modulus $m-M=19.5$, we obtain
the total luminosity of Draco in the $V$-band of $L_{\rm tot, V} =2.2 \times 10^5 L_{\sun}$,
which is somewhat larger than the earlier estimate of
$L_{\rm tot, V} =1.8 \times 10^5 L_{\sun}$ by Irwin \& Hatzidimitriou (1995).
We note that the adopted distance agrees well with the most recent estimate by
Bonanos et al. (2004) who found $m-M=19.4$ once an appropriate (for their
method of distance scale calibration) correction of 0.1 mag is applied.

In order to estimate the mass in stars, we need to adopt the stellar mass-to-light ratio
in $V$-band, which for dwarf spheroidals
is believed to be similar to that of globular clusters
$\Upsilon_V=(1-3) M_{\sun}/L_{\sun}$. Although the age and metallicity of Draco stellar
population are reasonably well known, the predicted values of $\Upsilon_V$ are still
uncertain and different stellar population synthesis codes and different initial mass
functions tend to produce results discrepant at least by a factor of 2 (Schulz et al.
2002). Since we do not want to overestimate the amount of dark matter in our modelling,
and keeping in mind that there is probably a significant contribution from ionized
gas in Draco (up to $2 \times 10^5 M_{\sun}$, Gallagher et al. 2003)
we take a higher value $\Upsilon_V=3 M_{\sun}/L_{\sun}$ so that our total stellar mass
will be $M_{\rm S} = \Upsilon_V L_{\rm tot, V} = 6.6 \times 10^5 M_{\sun}$. All the
parameters of Draco discussed in this Section are summarized in Table~\ref{draco}.

\begin{table}
\caption{Adopted parameters of the Draco dwarf. }
\label{draco}
\begin{center}
\begin{tabular}{ll}
parameter & value \\
\hline
luminosity $L_{\rm tot, V}$              & $2.2 \times 10^5 L_{\sun}$ \\
distance $d$                             & 80 kpc \\
distance modulus $m-M$                   & 19.5 \\
stellar mass-to-light ratio $\Upsilon_V$ &  $3 M_{\sun}/L_{\sun}$  \\
stellar mass $M_{\rm S}$                 & $6.6 \times 10^5 M_{\sun}$ \\
S\'ersic radius $R_{\rm S}$              & 7.3 arcmin  \\
S\'ersic parameter  $m$                  & 0.83  \\
\hline
\end{tabular}
\end{center}
\end{table}

We also note that the ellipticity of the surface density distribution
of stars in Draco is low, $e=0.3$ (Odenkirchen et al. 2001; Piatek et al. 2002),
so the galaxy is likely to be close to
spherical, which makes the use of our spherical models justified and reduces
the chance that non-sphericity will affect our results (see Sanchis et al. 2004).

\subsection{Dark matter}

Recently, Kazantzidis et al. (2004) performed a series of $N$-body simulations
of the evolution of a subhalo in the field of a larger halo of a galaxy. Their
Draco-like object initially had an NFW density distribution (Navarro, Frenk \& White
1997) but due to tidal interactions with a galaxy similar to the Milky Way lost
some of its mass. The resulting, tidally stripped, density profile retained the
initial $r^{-1}$ cusp at the centre but developed an exponential cut-off at large
distances and could be well approximated by a formula
\begin{equation}    \label{k1}
    \rho_{\rm d}(r) = C r^{-1} {\rm exp} \left( -\frac{r}{r_{\rm b}} \right),
\end{equation}
where $r_{\rm b}$ is the break radius at which the cut-off occurs. The dark
mass distribution associated with the density profile (\ref{k1}) is
\begin{equation}    \label{k2}
    M_{\rm d}(r) = M_{\rm D}
    \left[1 - {\rm exp} \left( -\frac{r}{r_{\rm b}} \right)
    \left(1+\frac{r}{r_{\rm b}} \right) \right] ,
\end{equation}
which we have normalized with $M_{\rm D}$, the total dark mass of the halo. (Note that
contrary to the standard NFW distribution, for this profile the mass converges.) With this
normalization, the constant in equation (\ref{k1}) becomes
$C=M_{\rm D}/(4 \pi r^2_{\rm b})$.

It is still debated whether tidal interaction of dwarfs with gravitational potential of
their host galaxies is likely to flatten the inner profile of their halo. Although
Kazantzidis et al. (2004) find that the cusp is preserved over a few orbital periods, there are
also claims that it can be flattened (Hayashi et al. 2003). One may also question the
assumption of a cuspy initial profile in the light of recent refinements in studies
of density profiles of dark matter haloes which find a flattening profile in the centre
(Navarro et al. 2004; Stoehr 2005). We believe however that,
from the point of view of the present study, the inner slope of the
halo is not really important since it should not affect our conclusions concerning
the total mass or anisotropy. In the study of the dark matter distribution in the Coma
cluster {\L}okas \& Mamon (2003) considered a generalized profile with an arbitrary inner slope
and found almost the same best-fitting mass and anisotropy for flat and cuspy inner profiles.

\section{Results and discussion}

In this Section, we report our attempts to reproduce the observed velocity moments
shown in Fig.~\ref{dispkurt} using models described
by equations (\ref{proj1})-(\ref{kurt}), with the mass distribution given by the sum of
the two contributions (\ref{m5d}) and (\ref{k2}) discussed in the previous Section:
\begin{equation}    \label{c7}
	M(r) = M_{*}(r) + M_{\rm d}(r) \ .
\end{equation}
The density profile $\nu(r)$ of the tracer population of stars is given by equation
(\ref{m5c}) and the surface brightness $I(R)$ by the S\'ersic formula (\ref{m5}) with
$m=0.83$.
As discussed by \L okas \& Mamon (2003) and Sanchis et al. (2004),
while studying velocity dispersion is useful
to constrain the mass, the kurtosis is mostly sensitive to the velocity anisotropy.
However, as will become clear below, when the anisotropy is constrained, the degeneracy
between the mass distribution and velocity anisotropy, usually present when
only velocity dispersion is considered, is broken and the parameters describing the
mass distribution can also be constrained.

\begin{figure}
\begin{center}
    \leavevmode
    \epsfxsize=7.8cm
    \epsfbox[45 40 320 790]{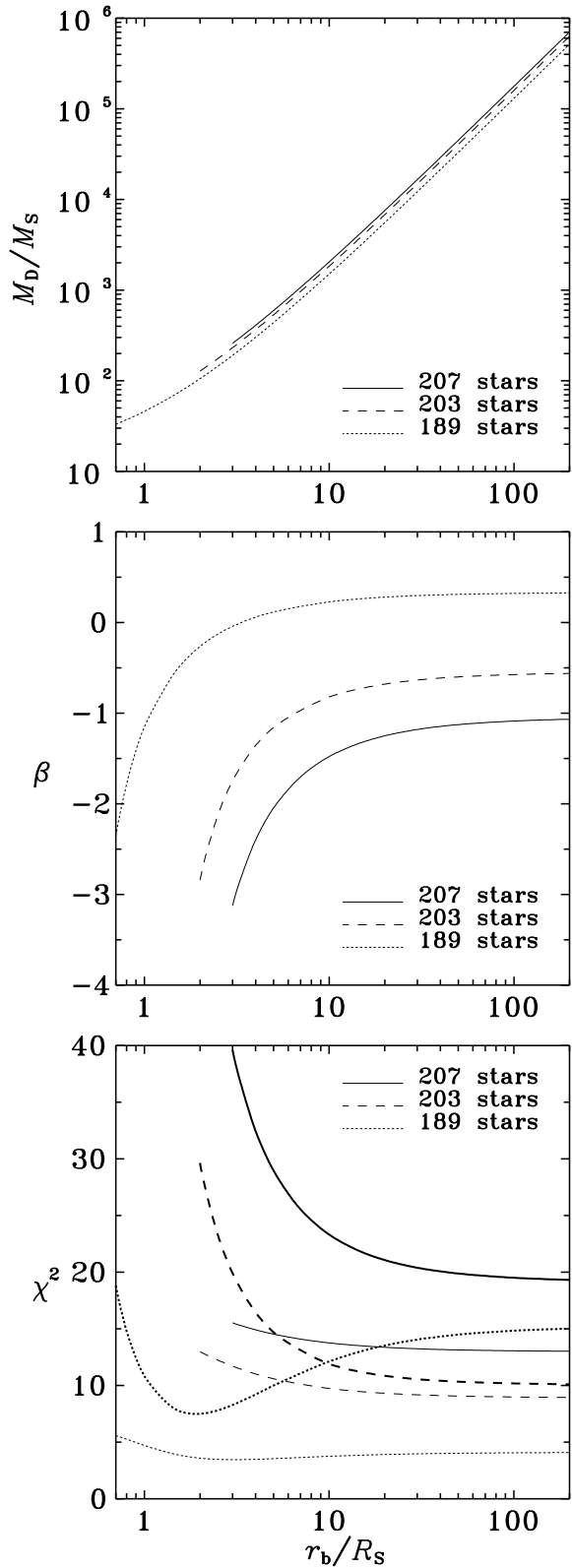}
\end{center}
\caption{Results of fitting only the line-of-sight velocity dispersion data.
The best fitting parameters $M_{\rm D}/M_{\rm S}$ and $\beta$ are shown in the two upper
panels as a function of $r_{\rm b}/R_{\rm S}$ for different stellar samples. The thinner
lines in the lower panel
give the goodness of fit $\chi^2$ from fitting velocity dispersion. The thicker lines
show
$\chi^2$ values obtained for the same parameters if the kurtosis data are included.}
\label{dispfit}
\end{figure}

We begin by fitting the line-of-sight velocity dispersion profiles shown by the upper
curves in Fig.~\ref{dispkurt} and adjusting three parameters:
the anisotropy parameter $\beta$, total dark mass in units of the total stellar mass
$M_{\rm D}/M_{\rm S}$ and the break radius of the dark matter profile in units
of the S\'ersic radius $r_{\rm b}/R_{\rm S}$ measuring the extent of the dark matter halo.
The results for our different samples of 189, 203 and 207 stars are shown in
Fig.~\ref{dispfit} so that the best-fitting $M_{\rm D}/M_{\rm S}$ (upper panel)
and the velocity anisotropy parameter
$\beta$ (middle panel) are displayed as a function of $r_{\rm b}/R_{\rm S}$.
The thinner lines in the lower panel of Fig.~\ref{dispfit} show the corresponding
goodness of fit $\chi^2$.
The lower panel of Fig.~\ref{dispfit} proves that none of the parameters
can be constrained from the analysis of velocity dispersion alone:
$\chi^2$ is flat for a large range of $r_{\rm b}/R_{\rm S}$ and
does not discriminate between different values of the parameters.

As discussed in Section~3, the sampling distributions of $\sigma_{\rm los}$
and $(\log \kappa_{\rm los})^{1/10}$ are independent to a good approximation,
hence we can use the same $\chi^2$
minimization scheme to find joint constraints following from
fitting both quantities. Before that, however, we calculate $\chi^2$
using the total of $10$ data points for each sample of
189, 203 and 207 stars for the values of the parameters dictated
by the best fit to the velocity dispersion data in
Fig.~\ref{dispfit}. The results are shown in the lower panel of the Figure with thicker
lines. We immediately see that $\chi^2$ has a minimum only for our sample with 189 stars.
For the other samples its values decrease for increasing $r_{\rm b}/R_{\rm S}$ and
$M_{\rm D}/M_{\rm S}$ and the minimum is not reached even if we go to
incredibly high masses like $M_{\rm D}/M_{\rm S} \approx 10^6$.
Such masses would be comparable to the mass of the Milky Way itself.

\begin{figure}
\begin{center}
    \leavevmode
    \epsfxsize=8cm
    \epsfbox[35 40 320 540]{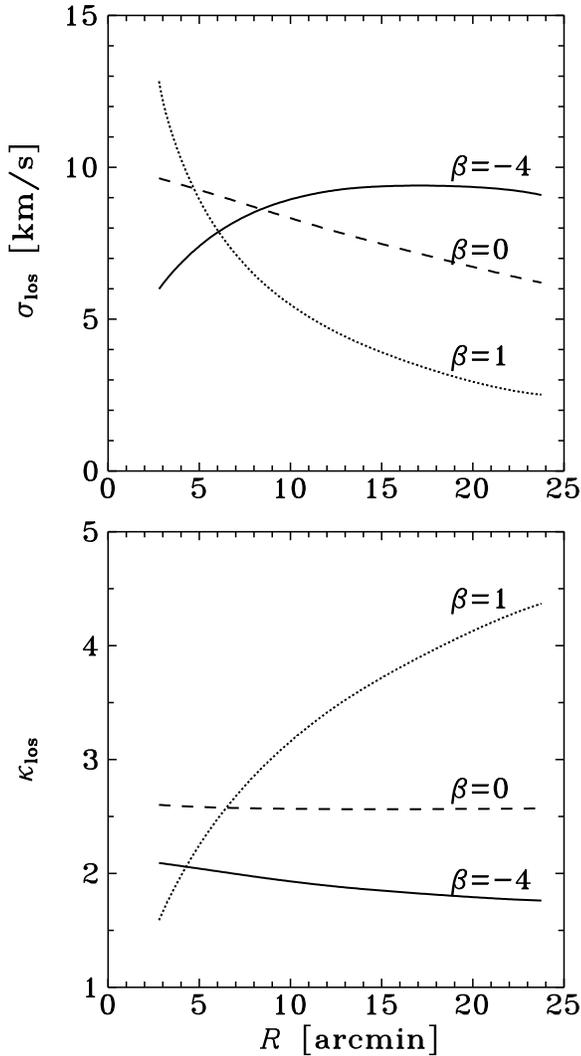}
\end{center}
    \caption{Line-of-sight velocity dispersion $\sigma_{\rm los}$ (upper panel)
	and kurtosis
	(lower panel) as a function of projected radius, for different
	anisotropies: radial ($\beta=1$), isotropic ($\beta=0$) and
	moderately tangential ($\beta = -4$, i.e. $\sigma_\theta/\sigma_r = 2.2$).
	The mass distribution is given by the best-fit for
	the sample with 189 stars, eq.~(\ref{parameters}).}
\label{dkpred}
\end{figure}

\begin{figure}
\begin{center}
    \leavevmode
    \epsfxsize=8cm
    \epsfbox[35 40 320 540]{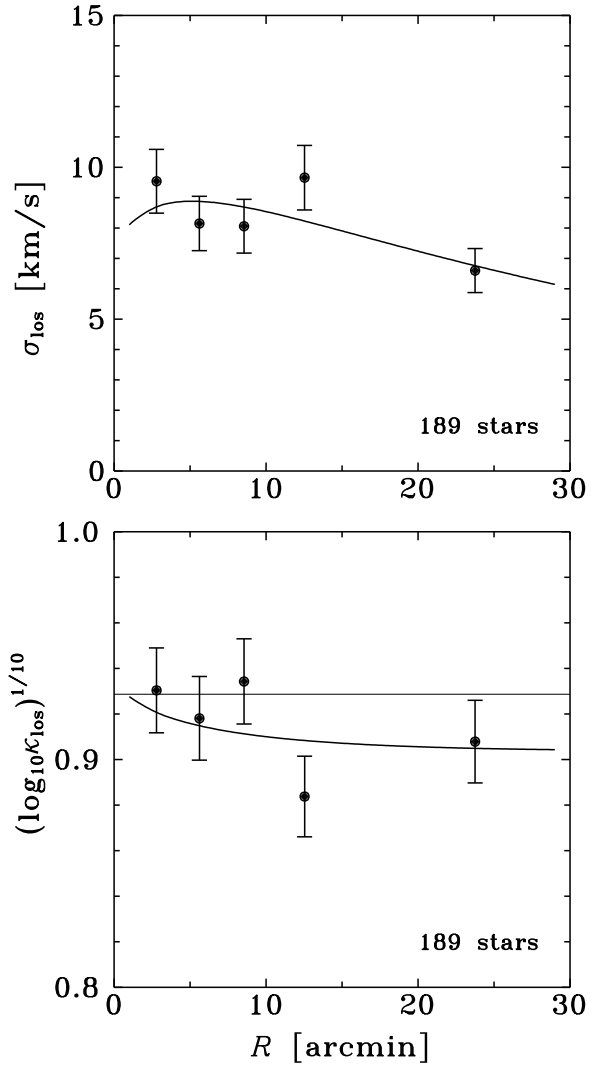}
\end{center}
    \caption{Line-of-sight velocity dispersion $\sigma_{\rm los}$ (upper panel)
	and dimensionless line-of-sight kurtosis parameter
	$(\log \kappa_{\rm los})^{1/10}$ (lower panel) with $1\sigma$ error bars for
        the sample with 189 stars.
	The curves represent the best-fitting model. The thin horizontal line in the
        lower panel
	marks the value of $(\log 3)^{1/10} = 0.93$ corresponding to the Gaussian velocity
	distribution.}
\label{dk189}
\end{figure}

The reason for this behaviour can be understood by referring to Fig.~\ref{dkpred}, where we plot
the predicted velocity moments for different values of the anisotropy parameter $\beta$.
For simplicity we assumed the mass distribution as obtained from the fit to the sample
of 189 stars (see eq.~[\ref{parameters}] below). It should be kept in mind that the kurtosis
profile depends mainly on anisotropy, while the velocity dispersion is degenerate in
anisotropy versus mass (taking a more extended mass distribution for the plot would result in
a more strongly increasing $\sigma_{\rm los}$ in analogy to more tangential orbits but the
kurtosis profile would remain unaltered).
As seen in the Figure and already discussed by
\L okas \& Mamon (2003) and Sanchis et al. (2004) (see also Gerhard 1993),
tangential orbits produce a decreasing
kurtosis profile, as is indeed observed for the sample with 189 stars. But the kurtosis
values of samples with 207 and 203 stars are rather high (close to Gaussian) and therefore
$\beta$ values close to isotropic or mildly radial are preferred while increasing
velocity dispersion profiles require tangential orbits (see the middle panel of
Fig.~\ref{dispfit}). Without tangential orbits
such velocity dispersion profiles could only be reproduced if the mass profiles are
extended and this is the reason the $\chi^2$ values decrease for higher masses and
break radii. This indicates that the samples with 207 and 203 stars should be treated
with caution. In the case of the sample with 203 stars a reasonably good fit to
both moments can be obtained, although for rather high masses. For the sample with 207
stars however, the fit is bad and improves only for incredibly high masses, which may
indicate that this sample yields inconsistent velocity dispersion and kurtosis profiles
and therefore that it almost certainly includes unbound stars.

One may ask to what extent this conclusion depends on our assumed constant anisotropy.
It has recently been verified using the cosmological $N$-body simulations
(Wojtak et al. 2005) that the radial velocity moments of bound objects,
eqs.~(\ref{sol1}) and (\ref{sol2}),
are actually sensitive to the local value of the anisotropy parameter. We expect
this to be true also for the projected moments since the dominant contribution to the projected
moment near the centre of the object comes from the stars that are actually near the
centre because they are more numerous than the foreground and background stars, while
the value of the projected moment at some projected distance $R$ from the centre is affected only
by stars located at true distances $r \ge R$. This behaviour in the case of velocity dispersion
has been explicitly verified by {\L}okas \& Mamon (2001) for the Osipkov-Merritt anisotropy
(see their Figures 1 and 6). We therefore expect that the anisotropy parameter changing e.g. from
isotropic orbits near the centre to tangential orbits at larger distances would produce
$\sigma_{\rm los}$ and $\kappa_{\rm los}$ profiles following the dashed line at small $R$ and
solid line at large $R$ in the upper and lower panel of Fig.~\ref{dkpred} respectively. The kurtosis
profile would then decrease even more strongly and the conclusion concerning the samples with
203 and 207 stars would be similar as in the case of constant $\beta$. Interestingly, the dispersion
profile interpolating between the dashed and solid line in the upper panel of Fig.~\ref{dkpred}
could reproduce the dip in the velocity dispersion profile seen in the data for all samples (see
Fig.~\ref{dispkurt}) and preserved for different binnings of the stars.

We therefore conclude that the choice of the sample of stars is the dominant source of
uncertainty when trying to determine the mass distribution in Draco.
In the following we will only consider the sample with 189 stars and investigate the
next important source of uncertainty following from the sampling errors we assigned to our data
points. The errors in the adopted parameters of Draco concerning the
surface distribution of stars, luminosity or distance probably have much
smaller effect (see {\L}okas 2002).

The joint fitting of the velocity dispersion and kurtosis for the sample with 189 stars
gives results not very different from those indicated by the minimum of $\chi^2$ (thicker
dotted line) in the lowest panel of Fig.~\ref{dispfit}.
The best-fitting parameters (with $\chi^2/N=7.4/10$) are
\begin{equation}    \label{parameters}
	M_{\rm D}/M_{\rm S} = 99, \ \
	r_{\rm b}/R_{\rm S} = 1.9, \ \
	\beta = -0.4.
\end{equation}
The best estimate of the total mass of the Draco dwarf is then $6.6 \times 10^7 M_{\sun}$.
We emphasize that the result is not robust and could be very different if
different stellar samples were
taken. Our choice of velocities has been rather arbitrary and one could probably get
reasonable results also with slightly larger or smaller samples. One could for example
construct samples by removing the stars with most discrepant velocities one by one and
performing a similar fitting
procedure as described here. One would then get a series of best fitting masses of
decreasing values.

The mass estimate found above is significantly smaller than the ones obtained with the
NFW mass distribution and older data (\L okas 2002), which were even of the order of
$ 10^9 M_{\sun}$ for reasonable concentrations. This was due to the more extended
nature of the NFW profile, which was modified here by adding an exponential cut-off.
Recently, it has been proposed that, if the dwarf spheroidals of the Local Group indeed
possess such high masses, then their low abundance could be understood by referring to
the cosmological mass function, which predicts many fewer objects with such high mass
(Hayashi et al. 2003).
Our present lower mass estimate could be consistent with this solution to the
overabundance problem if the dwarfs originally indeed had such high masses but were then
tidally stripped in the Milky Way potential. Such a scenario could also explain why
dwarfs managed to build up their stellar content in presently shallow potential wells
(Kravtsov, Gnedin \& Klypin 2004).

Our estimate of Draco's mass is not directly comparable to that of
Kleyna et al. (2002) since they used
different dark matter models and a smaller stellar sample (constructed from
an older data set but in a weakly restrictive way similar to our sample with 207 stars).
Their best estimate
of the mass inside three core radii was $8 \times 10^7 M_{\sun}$, which is,
as would be expected, already more than our total mass value.

\begin{figure}
\begin{center}
    \leavevmode
    \epsfxsize=8cm
    \epsfbox[40 40 320 300]{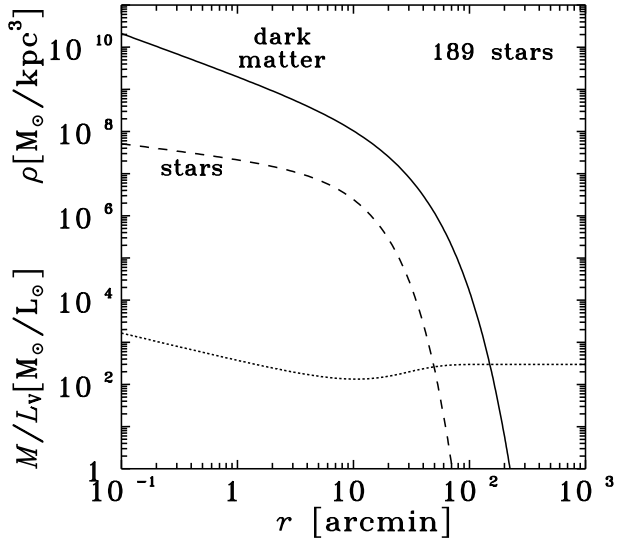}
\end{center}
\caption{
Best fitting dark matter density profile (solid line) in comparison to the stellar
mass density profile (dashed line). The dotted line shows the cumulative
mass-to-light ratio for the best fitting profile.}
\label{denmtl}
\end{figure}

The velocity moments obtained with the sets of parameters listed in (\ref{parameters})
are shown in Fig.~\ref{dk189} together with the data for 189 stars.
The dark matter density profile following from equation (\ref{k1}) with those parameters
is plotted in Fig.~\ref{denmtl} together with the stellar mass
density profile $\rho_{*} (r) = \Upsilon \nu(r)$ with $\nu(r)$ given by equation
(\ref{m5c}). Fig.~\ref{denmtl} also shows the cumulative mass-to-light
ratio, i.e. the ratio of the total
mass distribution to the luminosity distribution in the $V$-band inside radius $r$
\begin{equation}    \label{m2lratio}
	M/L_{\rm V} = \frac{M(r)}{L_{\rm V}(r)} ,
\end{equation}
where $M(r)$ is the sum of two contributions from stars and dark matter
given by equation (\ref{c7}) with (\ref{m5d}) and (\ref{k2}),
while $L_{\rm V}(r) = M_{*}(r)/\Upsilon_{\rm V}$.
The behaviour of this quantity is easily understood:
the dark matter cusp is steeper than that of the luminosity distribution therefore the
ratio increases towards the centre. This is due to our adopted profile of dark matter halo,
which, according to the simulations of Kazantzidis et al. (2004), preserves
the NFW cusp. With a flatter dark matter inner slope
(as recently proposed by Navarro et al. 2004 and Stoehr 2005 for global
halos, and by Hayashi et al. 2003 for tidally stripped ones), the data may well be
reconciled with a constant mass-to-light ratio.

At large distances, the mass-to-light ratio converges to a constant
value $\Upsilon_{\rm V} (1+M_{\rm D}/M_{\rm S})=299$
since both luminosity and dark matter distributions fade exponentially. This result is comparable
to the estimate by Kleyna et al. (2002) who find $M/L_{\rm V} = 330 $ within three core radii, while
Odenkirchen et al. (2001), assuming that mass follows light, obtain $M/L_{\rm i} = 146$. Our result
is also consistent with the study of {\L}okas (2002) using older data, where dark matter was
modelled with a generalized NFW distribution. There $M/L_{\rm V} > 100 $ was
found at $R>10$ arcmin, and the mass-to-light ratio was increasing at large radii due to
the assumption of an untruncated, $\rho \propto r^{-3}$, dark matter envelope.

\begin{figure}
\begin{center}
    \leavevmode
    \epsfxsize=7.7cm
    \epsfbox[95 70 295 640]{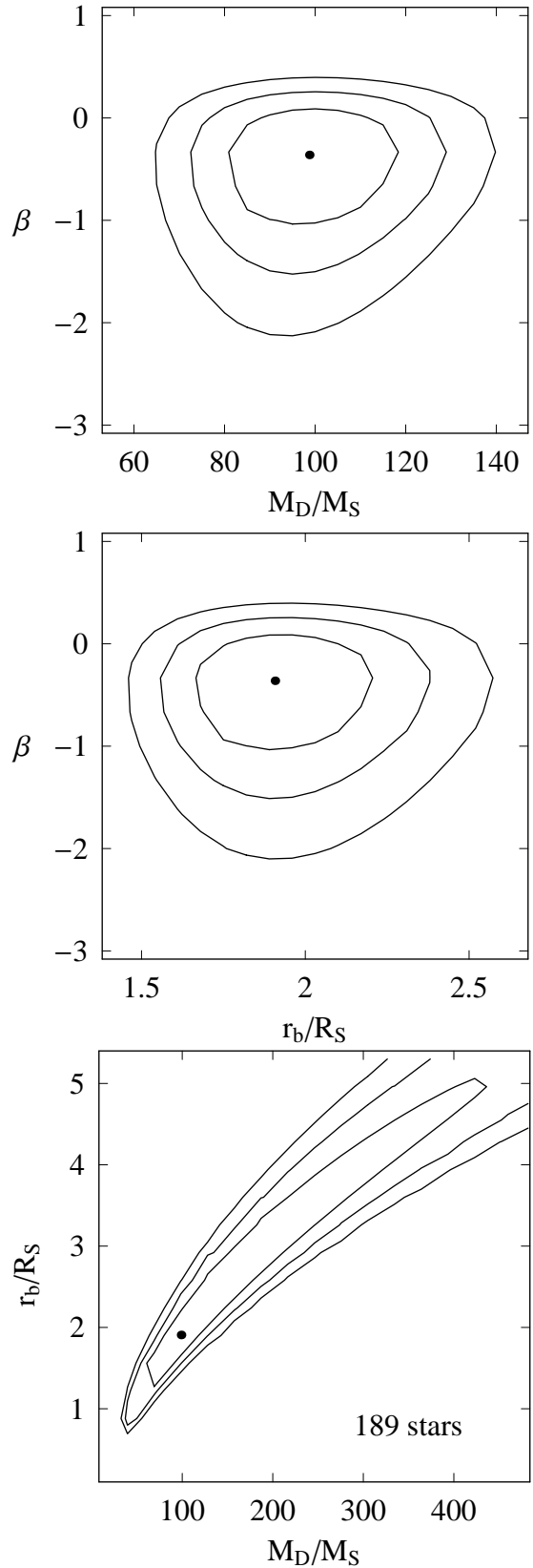}
\end{center}
\caption{Cuts through the $1\,\sigma$, $2\,\sigma$ and $3\,\sigma$
probability contours in
the parameter space obtained from fitting $\sigma_{\rm los}$
and $(\log \kappa_{\rm los})^{1/10}$ for the sample with 189 stars.
Dots indicate the best-fitting parameters.}
\label{contours}
\end{figure}

Fig.~\ref{contours} illustrates the uncertainties in the estimated
parameters due to the sampling errors of the velocity moments.
Here, we plot the cuts through the confidence region in three possible parameter planes
with probability contours corresponding
to $1\sigma$ ($68$ percent confidence), $2\sigma$ ($95$ percent) and $3\sigma$
($99.7$ percent) i.e.
$\Delta \chi^2 = \chi^2 - \chi^2_{\rm min} = 3.53, 8.02, 14.2$, where
$\chi^2_{\rm min}=7.4$.
Contrary to the case
where only the velocity dispersion was fitted, some constraints on parameters can now be
obtained, but the allowed ranges of parameters are large. This is especially the case for
parameters $M_{\rm D}/M_{\rm S}$ and $r_{\rm b}/R_{\rm S}$ which are highly
degenerate.
Since the parameters $M_{\rm D}/M_{\rm S}$ and $r_{\rm b}/R_{\rm S}$ are very
weakly correlated
with $\beta$, as can be seen from the two upper panels of Fig.~\ref{contours}, the cuts
provide a good approximation of
the true uncertainties in the parameters. However, these errors, due to the
low number of velocities per bin, are much smaller than uncertainties due to our choice of
stars to include in the sample.

As already stated above, we find the preferred stellar orbits to be weakly tangential. For the
sample with 189 stars we get the best-fitting $\beta=-0.4$ (corresponding to
$\sigma_\theta/\sigma_r \approx 1.2$) and the expected uncertainties can be read from
the two upper panels of Fig.~\ref{contours}. Note that according to the Figure the orbits
are consistent with isotropy at $1\,\sigma$ level, while
a strongly radial or a strongly tangential velocity distribution are ruled out.
It must be remembered, however, that these conclusions were reached with the assumption
of negligible binary fraction in Draco. As discussed in Section~3,
the correction for binaries
significantly affects the values of kurtosis (see Fig.~\ref{dispkurt}),
but again more so for the sample with 189
stars than for the other samples which have higher values of kurtosis.
If the binary fraction in Draco is significant, e.g.
of the order of $\gamma=0.5$ as for the solar neighbourhood according to
current estimates (Duquennoy \& Mayor 1991; De Rijcke \& Dejonghe 2002)
then the intrinsic kurtosis will be decreased and the inferred orbits even more tangential.
Reproducing the velocity dispersion profile (which is not strongly affected by the
binaries) would then require a less extended mass distribution and the inferred masses
will be even smaller.

We have mentioned before that the main gain from including the kurtosis in the analysis is the
ability to constrain the anisotropy of stellar orbits.
Interestingly, weakly tangential orbits in Draco appear to be a robust conclusion from our study;
whatever the sample of the stars taken and binary fraction assumed the preferred models
have $\beta<0$ or $\sigma_\theta/\sigma_r >1$. This result seems to contradict the
general belief, based e.g. on $N$-body simulations, that virialized systems
should have orbits close to isotropic or even mildly radial (see Mamon \& \L okas 2005).
However, as discussed by Kazantzidis et al. (2004),
tidally stirred stellar populations can develop tangential orbits even when evolving in
an isotropic halo.

A possible scenario for the formation of dwarf spheroidal galaxies
has been proposed by Mayer et al. (2001) who used high-resolution $N$-body/SPH
simulations to study the evolution of rotationally supported dwarf irregular galaxies
moving on bound orbits in the massive dark matter halo of the Milky Way. They showed
that the tidal field induces severe mass loss in their halos and the
low surface brightness dwarfs are transformed into pressure-supported dwarf spheroidals
with preferably tangential orbits outside the centre (see Fig. 25 of the preprint
version of Mayer et al. 2001).

Besides,
tangential orbits may not be restricted to dwarf spheroidals. A recent study of kinematics
of a sample of Virgo Cluster dwarf ellipticals (Geha, Guhathakurta \& van der Marel 2002)
also shows the preference of tangential orbits and actually the mechanism to produce them
may be similar as in the case of dwarf spheroidals (Mayer et al. 2001)
only with high surface brightness galaxies as progenitors.
Also, elliptical galaxies formed by major mergers of spiral galaxies often
show tangential anisotropy in their inner regions (Dekel et al. 2005), which
is caused by some of the stars having formed within the gaseous disk, which
is in nearly circular rotation in the merger remnant.

\section{Interlopers from the Milky Way or tidal debris?}

\begin{figure}
\begin{center}
    \leavevmode
    \epsfxsize=8cm
    \epsfbox[20 160 580 680]{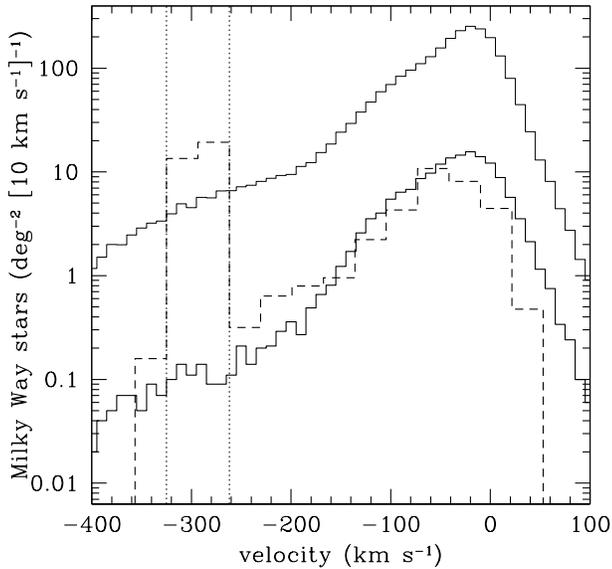}
\end{center}
\caption{
Expected contamination of Draco by Milky Way stars. The diagram shows the expected
total number of Milky Way stars of given velocities (higher histogram, thin solid line) and the
corresponding number if we restrict ourselves to the red giant branch
of the color-magnitude diagram of Draco (lower histogram, thick solid line). The dashed line
shows the velocity distribution of stars actually observed in the direction of Draco by Wilkinson
et al. (2004).
The contamination by giants in the velocity range of Draco, indicated by two vertical
dashed lines, is very small.}
\label{vcounts2}
\end{figure}

We have shown that the main issue in the analysis of the velocity moments is the question
of which stars should be included in the analysis, or, to put it in a different
way, which of them are unbound and only artificially affect the moments, e.g. inflate
the velocity dispersion. We argued that the presence of unbound stars leads to inconsistent
velocity dispersion and kurtosis which cannot both be fitted by equilibrium models
following from the Jeans formalism. It is therefore essential to study the
possible origin of those unbound stars.

An immediate first guess would be that the stars are the members
of the Milky Way. In order to verify this hypothesis we have ran the Besancon
Galaxy model (Robin et al. 2003, http://bison.obs-besancon.fr/modele/) in the direction of Draco,
which gives the velocity distribution shown by the higher histogram (thinner solid line) in
Fig.~\ref{vcounts2}. If stars of all types are considered, as they are for this histogram,
then the expected number of Milky Way interlopers with Draco-like velocities
in the range $-325$ km s$^{-1} < v < -262$ km s$^{-1}$, indicated by the two vertical
dashed lines in the Figure, would be $15 \pm 0.25$. However, if we restrict the analysis
to the red giant branch (RGB) of the Draco color-magnitude diagram
with $\pm 0.075$ mag from the typical trend we end up with a much lower expected number
density of stars represented by the lower
histogram (thicker solid line) in Fig.~\ref{vcounts2}. Since the observations of Draco stars
by Kleyna et al. (2002) and Wilkinson et al. (2004) were indeed restricted to RGB stars
we expect the contamination from the Milky Way stars to be very small with $1.42 \pm 0.17$
expected interlopers. The velocity distribution of all RGB stars observed by Wilkinson et al.
(2004) is shown with a dashed line in the plot. We can see that the level of this histogram right
outside the range of velocities adopted for Draco stars is somewhat higher than expected from the
Besancon model.
A na\"{\i}ve interpolation of the observed velocity distribution (dashed
histogram in Fig.~\ref{vcounts2}) outside of the Draco velocity zone would
yield $6.5\pm2.3$
interlopers from the Milky Way.

Given our finding that the kinematical modelling of the
189-star sample yields consistent  results, while those of the 207- and 203-star samples
do not, the total number of interlopers in Draco is of order of $207-189=18$ stars.
Therefore, given our estimates of the previous paragraph of the number of
Milky Way interlopers, an important fraction (probably
the majority) of unbound stars in the neighbourhood of Draco dwarf must
originate in Draco
itself.
An obvious mechanism for the production of such stars is the tidal stripping of
Draco by the Milky Way potential. Although no clear signature of tidal tails has been
identified in the luminosity distribution of Draco (e.g. Klessen et al. 2003),
it is still quite possible that
the tails are oriented along the line of sight.

Wilkinson et al. (2004) have recently
argued that the tidal scenario is only feasible if the mass of Draco is below $5 \times 10^7
M_{\sun}$ and its dark matter distribution is low outside roughly 30 arcmin. Interestingly,
the results of our analysis seem to almost fulfill these requirements: our best estimate of
the mass for the sample with 189 stars is $6.6 \times 10^7
M_{\sun}$ and the dark matter distribution indeed fades at larger distances, as
demonstrated by Fig.~\ref{denmtl}. Wilkinson et al. (2004) also point out that such a tidally
detached population of stars
should be heated up, which is contradicted by the decreasing velocity dispersion of Draco
at large distances, unless only cold stars remain which are then recaptured by the dwarf.
We find that the fall of the velocity dispersion at large angular distances from the
centre of Draco is not a very strong feature and its visibility depends on
the chosen radial binning
of stars.

\section{Concluding remarks}

Given the new
data on the Draco dwarf spheroidal,
especially the velocity measurements by Kleyna
et al. (2002) and Wilkinson et al. (2004),
we have attempted to constrain the dark matter
distribution in Draco by modelling not only the dispersion,
but also
the kurtosis of the line-of-sight velocity distribution. The analysis of both moments
allows us to break the degeneracy between the mass distribution and velocity anisotropy
usually present in the analyses of velocity dispersion. Still, many uncertainties remain
in the analysis. We show that the results can
be very different depending on the sample of stars chosen, i.e. which
stars are included in the sample and which are treated as interlopers.
Besides, due to the limited number
of measured velocities, the sampling errors of the moments are large and the parameters
cannot be strongly constrained.

There are few possible remedies to improve the present situation: to have more
velocity measurements, to further constrain the dark matter profiles on
theoretical grounds and to model the influence of interlopers in more detail.
All these approaches seem feasible in the near future. On the
observational side, the measurements of a few hundred velocities per dwarf galaxy
are planned or already being performed using large telescopes with multi-fiber spectroscopy like
Magellan (M. Mateo, private communication). On the theoretical side, progress
might be even faster with the rapidly increasing resolution of the $N$-body simulations.
The recent result of Kazantzidis et al. (2004) itself which we have used in the present
work relied on increased resolution to show that the cusp of the dark matter profile
is not flattened due to tidal interaction as was claimed earlier (Stoehr et al. 2002; Hayashi
et al. 2003).
The result however depended to some extent on the parameters of the orbit of the dwarf
around the Milky Way, which in the case of Draco are poorly known.

As for the third issue, the treatment of
interlopers or unbound stars, in our opinion this is at present
the most important problem in the analysis of the
dark matter distribution in dwarf spheroidal galaxies. We have shown that the supposed
interlopers can alter the profiles of velocity moments and significantly change the
estimated parameters of the dark matter distribution.


\section*{Acknowledgements}

We are grateful to M. Wilkinson and collaborators for sharing with us their new velocity
measurements for Draco prior to publication.
We wish to thank E. Grebel, J. Ka{\l}u\.{z}ny, S. Kazantzidis,
J. Kleyna, A. Klypin, J. Miko{\l}ajewska, A. Olech,
K. Stanek, F. Stoehr, and especially M. Wilkinson for
discussions and comments,
and an anonymous referee for suggestions which helped to improve the paper.
E{\L} is grateful for the hospitality of Institut d'Astrophysique de Paris and
Instituto de Astrof{\'\i}sica de Andalucia
in Granada where part of this work was done.
We made use of the Besancon Galaxy model available at
http://bison.obs-besancon.fr/modele/.
This research was partially supported by the
Polish Ministry of Scientific Research and Information Technology
under grant 1P03D02726,
the Jumelage program Astronomie France Pologne of CNRS/PAN and the exchange program of CSIC/PAN.


\vspace{-0.2in}

\begin{thebibliography}{}

\bibitem[]{acm} Aparicio A., Carrera R., Martinez-Delgado D., 2001, AJ, 122, 2524
\bibitem[]{bt} Binney J., Tremaine S., 1987, Galactic Dynamics. Princeton
    Univ. Press, Princeton, chap. 4.
\bibitem[]{bo} Bonanos A. Z., Stanek K. Z., Szentgyorgyi A. H., Sasselov D. D.,
	Bakos G. A., 2004, AJ, 127, 861
\bibitem[]{dd} De Rijcke S., Dejonghe H., 2002, MNRAS, 329, 829
\bibitem[]{dekel} Dekel A., Stoehr F., Mamon G. A., Cox T. J., Novak G.,
	Primack J. R., 2005, Nat, in press, astro-ph/0501622
\bibitem[]{dms} Diemand J., Moore B., Stadel J., 2004, MNRAS, 352, 535
\bibitem[]{dm} Duquennoy A., Mayor M., 1991, A\&A, 248, 485
\bibitem[]{wsi} Fukugita M., Shimasaku K., Ichikawa T., 1995, PASP, 107, 945
\bibitem[]{gall} Gallagher J. S., Madsen G. J., Reynolds R. J., Grebel E. K.,
	Smecker-Hane T. A., 2003, ApJ, 588, 326
\bibitem[]{geha} Geha M., Guhathakurta P., van der Marel R. P., 2002, AJ, 124, 3073
\bibitem[]{ger} Gerhard O. E., 1993, MNRAS, 265, 213
\bibitem[]{hayashi} Hayashi E., Navarro J. F., Taylor J. E., Stadel J., Quinn T., 2003,
	ApJ, 584, 541
\bibitem[]{ih} Irwin M., Hatzidimitriou D., 1995, MNRAS, 277, 1354
\bibitem[]{kmm} Kazantzidis S., Magorrian J., Moore B., 2004, ApJ, 601, 37
\bibitem[]{kmmds} Kazantzidis S., Mayer L., Mastropietro C., Diemand J., Stadel J.,
	Moore B., 2004, ApJ, 608, 663
\bibitem[]{kk} Klessen R. S., Kroupa P., 1998, ApJ, 498, 143
\bibitem[]{kz}  Klessen R. S., Zhao H., 2002, ApJ, 566, 838
\bibitem[]{kgh} Klessen R. S., Grebel E. K., Harbeck D., 2003, ApJ, 589, 798
\bibitem[]{kl1} Kleyna J. T., Wilkinson M. I., Evans N. W., Gilmore G., 2002,
	MNRAS, 330, 792
\bibitem[]{kkvp} Klypin A., Kravtsov A. V., Valenzuela O., Prada F., 1999, ApJ, 522, 82
\bibitem[]{ln} Lima Neto G. B., Gerbal D., M\'arquez I., 1999, MNRAS, 309, 481
\bibitem[]{lok} \L okas E. L., 2001, MNRAS,  327, 21P
\bibitem[]{lo} {\L}okas E. L., 2002, MNRAS, 333, 697
\bibitem[]{lm03} {\L}okas E. L., Mamon G. A., 2003, MNRAS, 343, 401
\bibitem[]{ml05} Mamon G. A., {\L}okas E. L., 2005, MNRAS, in press, astro-ph/0405491
\bibitem[]{m1} Mateo M., 1997, in Arnaboldi M. et al., eds,
	ASP Conf. Ser. Vol. 116, The Nature of Elliptical
	Galaxies. Astron. Soc. Pac., San Francisco, p. 259
\bibitem[]{mayer} Mayer L., Governato F., Colpi M., Moore B., Quinn T., Wadsley J.,
	Stadel J., Lake G., 2001, ApJ, 559, 754, a slightly different version at
	astro-ph/0103430
\bibitem[]{mk} Merrifield M. R., Kent S. M., 1990, AJ, 99, 1548
\bibitem[]{nfw97} Navarro J. F., Frenk C. S., White S. D. M., 1997, ApJ,
    490, 493 (NFW)
\bibitem[]{n04} Navarro J. F. et al., 2004, MNRAS, 349, 1039
\bibitem[]{oden} Odenkirchen M. et al., 2001, AJ, 122, 2538
\bibitem[]{ppao} Piatek S., Pryor C., Armandroff T. E., Olszewski E. W.,
	2002, AJ, 123, 2511
\bibitem[]{prada} Prada F. et al., 2003, ApJ, 598, 260
\bibitem[]{robin} Robin A. C., Reyl\'e C., Derri\`ere S., Picaud S., 2003,
	A\&A, 409, 523
\bibitem[]{slm} Sanchis T., \L okas E. L., Mamon G. A., 2004, MNRAS, 347, 1198
\bibitem[]{sfmf} Schulz J., Fritze-v. Alvensleben U., Moeller C. S., Fricke K. J.,
	2002, A\&A, 392, 1
\bibitem[]{s} S\'ersic J. L., 1968, Atlas de Galaxies Australes,
	Observatorio Astronomico, Cordoba
\bibitem[]{stoehr} Stoehr F., 2005, MNRAS, submitted, astro-ph/0403077
\bibitem[]{swts} Stoehr F., White S. D. M., Tormen G., Springel V., 2002, MNRAS, 335, 84P
\bibitem[]{vdm} van der Marel R. P., Magorrian J., Carlberg R. G., Yee H.
    K. C., Ellingson E., 2000, AJ, 119, 2038
\bibitem[]{wkeg} Wilkinson M. I., Kleyna J. T., Evans N. W., Gilmore G. F.,
	Irwin M. J., Grebel E. K., 2004, ApJ, 611, L21
\bibitem[]{wlgm} Wojtak R., {\L}okas E. L., Gottl\"ober S., Mamon G. A., 2005,
	MNRAS, 361, L1


\end{thebibliography}
\end{document}